# Waveguide integrated low noise NbTiN nanowire single-photon detectors with milli-Hz dark count rate

Carsten Schuck, Wolfram H. P. Pernice, Hong X. Tang*

*Department of Electrical Engineering, Yale University, New Haven, Connecticut, 06511, United States*

*E-mail: hong.tang@yale.edu.

Subject areas: Superconducting single photon detectors, nanophotonics, silicon nitride, niobium titanium nitride, quantum optics

Superconducting nanowire single-photon detectors are an ideal match for integrated quantum photonic circuits due to their high detection efficiency for telecom wavelength photons. Quantum optical technology also requires single-photon detection with low dark count rate and high timing accuracy. Here we present very low noise superconducting nanowire single-photon detectors based on NbTiN thin films patterned directly on top of $Si_3N_4$ waveguides. We systematically investigate a large variety of detector designs and characterize their detection noise performance. Milli-Hz dark count rates are demonstrated over the entire operating range of the nanowire detectors which also feature low timing jitter. The ultra-low dark count rate, in combination with the high detection efficiency inherent to our travelling wave detector geometry, gives rise to a measured noise equivalent power at the $10^{-20}$ $W/Hz^{1/2}$ level.

Low dark count rate, high detection efficiency and accurate timing resolution are the three most desired features of a single-photon detector (1,2). These characteristics can be combined to one figure-of-merit for single-photon detectors as for example the noise equivalent power (NEP). Detectors with low noise performances are increasingly sought after for applications in both quantum and classical technology. In particular linear optics quantum information processing crucially relies on the availability of low dark count rate single-photon detectors (1,3-5). Most prominently, quantum key distribution implementations (6,7) are currently limited in rate and range by imperfect detector characteristics. Other applications which will greatly benefit from improved single-photon detection systems include the characterization of quantum emitters (8-11), optical time domain reflectometry (12) as well as picosecond imaging circuit analysis (13).

One of the most promising technologies to achieve low noise detector characteristics are nanowire superconducting single-photon detectors (SSPDs) (14). The detection mechanism relies on single-photon induced hotspot creation in a superconducting nanowire which is current biased closed to its critical current (15). The detection process is characterized by a fast recovery time and high quantum efficiency both for visible and infrared wavelength photons (1,16,17). Until recently most state-of-the-art SSPDs however, were stand-alone units absorbing fiber-coupled photons in a superconducting niobium nitride (NbN) thin film under normal incidence which limits their usefulness for large scale integrated photon counting applications. Such difficulties can be overcome by patterning NbN nanowires on silicon waveguides to realize high on-chip quantum efficiency (18). However, the performance of NbN-SSPDs is limited by detection noise, where dark count rates can approach the kHz-range.

SSPD characteristics are fundamentally determined by substrate and superconducting material properties (15,19-21). High quantum efficiency is easier achieved in low energy gap superconductors (17) whereas NbTiN is a superior material choice in terms of low noise performance (22). Here we show that ultra-low dark count rates can be achieved with amorphous NbTiN-on-SiN nanowire single-photon detectors. Our results represent an improvement of several orders of magnitude over SSPDs made from NbN which typically only achieve low dark count rates ($R_{dc}$) when the bias current is set far below the critical current (3,18,23,24). Furthermore, we achieve highly efficient single-photon detection by engineering the detector and waveguide dimensions at the nanoscale. Our detectors thus feature very low NEP on a fully scalable platform, addressing the most urgent needs of integrated quantum optical information processing (18,25) and time correlated single-photon counting applications.

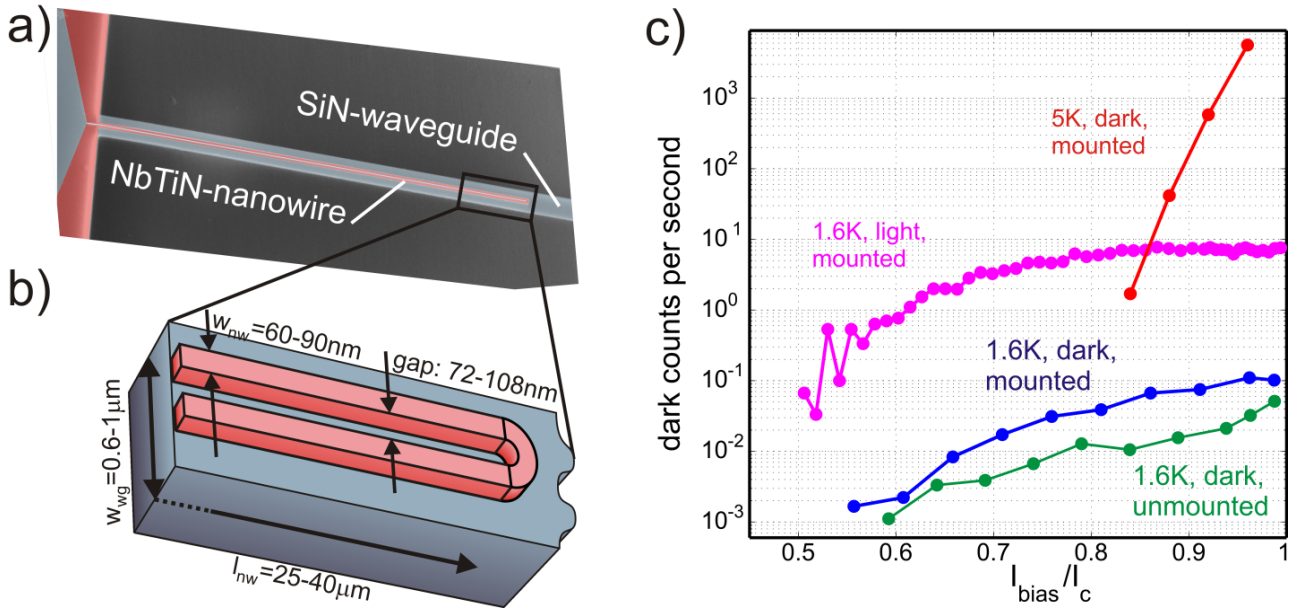

**Figure 1**. NbTiN-nanowire patterned directly on top of SiN waveguides; a) false-color scanning electron micrograph of a waveguide integrated SSPD; b) nanowire geometry with width $w_{nw}$ varied from 60 nm to 90 nm and length $l_{nw}$ varied between 25 μm and 40 μm. The nanowire is 8 nm thick. The gap between the wires is designed to be 120% of the respective nanowire width. The waveguides are 330 nm high at a width of $w_{wg}$=600 nm for visible light and 1 μm for telecom wavelength light; c) Dark count rate as a function of normalized bias current. To mimic realistic measurement conditions we align the fiber array to the on-chip waveguides ("mounted" detector) and cap the fiber input. The resulting dark count rate at minimal ambient light levels ("dark") is shown as blue trace at a temperature of 1.6 K and as a red trace at 5 K. When the ambient light level is increased to daylight conditions ("light") we observe the magenta curve (at 1.6 K). Displacing the fiber array from the chip reduces the stray light guided to the detector ("unmounted" detector) and we observe the green trace at 1.6 K under low ambient light levels.

## Results

**Low dark count rate NbTiN detectors.** We fabricate superconducting nanowire detectors from 8 nm thin NbTiN films deposited on SiN-on-insulator wafers (for details on the fabrication process see ref. (26)), as shown in figure 1 a) and b). To test these detectors under realistic conditions for practical detector applications we pattern wideband transparent waveguides out of the SiN-substrate layer to route visible and infrared photons with low loss to the nanowire-covered SSPD-region. Optical links to the SSPD-sample chip mounted inside a liquid helium cryostat are provided via a fiber array (see Methods).

We first perform dark count rate measurements during optimal alignment of the fiber array to the on-chip waveguides, as it would be the case during regular detector operation. Although the cryostat itself is well shielded, residual stray light does penetrate through the optical fiber into the cryostat and further to the detector via the waveguide ("mounted" detector). The resulting dark count rate for a 75 nm-wide NbTiN nanowire of 40 μm length is shown in figure 1 c) and was recorded at 1.6 K under "dark" laboratory conditions, i.e. all laboratory light sources are reduced to minimal illumination levels. Under these conditions we find a maximal dark count rate of 100 mHz at >96% of the critical current, which was 18 μA for this detector. The rate then gradually decreases to 2 mHz at 55% $I_c$.

To assess the intrinsic dark count rate of our SSPDs we also record the detector output pulses after displacing the fiber array from the on-chip waveguide couplers to avoid stray light being guided from the fiber to the (now "unmounted") detector. We observe a further decrease of the number of dark counts as compared to the above setting, reaching a maximal rate of 50 mHz at 99% of the critical current. In this case the lowest dark count rate of 1 mHz was recorded at 60% $I_c$ where the detection efficiency remains high (see Fig. 2). At even lower bias currents the dark count rate is so low, that the data acquisition time of 30 minutes per measurement point does not allow us to acquire enough counts to report meaningful statistics. Our NbTiN nanowire SSPDs thus achieve low intrinsic dark count rates over their entire bias current range.

In a third series of measurements we increase the ambient illumination level to daylight conditions and again record the dark count rate which now reaches a maximal value of 7.7 Hz confirming the sensitivity of our detectors to ultra-low light levels. For none of the three curves recorded at 1.6 K do we observe an exponential rise of the dark count rate at bias currents close the critical current as reported in previous studies (3,22-24). We attribute this to the fact that the dominant decoherence mechanisms of the superconducting state, e.g. unbinding of vortex-antivortex pairs and vortex hopping over the edge barrier (24,27), are suppressed below the stray light level of our apparatus in this temperature range for 8 nm thick NbTiN nanowires. Reducing stray light levels even further may allow us to study the origin of intrinsic dark count rates in nano-patterned NbTiN structures below 2 K with our setup.

To verify that bias-current assisted thermal fluctuations persist in our NbTiN nanowires we heat our sample to 5 K and measure the dark count rate when biasing the wire close to the critical current. As shown in Fig 3 c), an exponential dependence of the dark count rate on the bias current is indeed observed in this case. During this measurement ambient light levels were reduced to a minimum but the data acquisition time per point was too short to resolve the stray light background at lower bias current.

**Detector geometry optimization.** To simultaneously achieve low noise and high detection efficiency over a broad bias range we employ a traveling wave design (18,26,28,29), as shown in figure 1, and optimize the geometry of the nanowires. Because a relatively high percentage of the incoming photons will be absorbed at the 180 degree turn of the nanowire (where the light is first incident) we employ rounded bends at this point (see Fig. 1b)). When current flows around a bend it will crowd at the inner edge, leaving the outer side of the bend with less current density (30-32). In sharp-angled bend-geometries the current density is particularly low at outside corners. In contrast, our U-shaped wire geometry minimizes NbTiN areas which contribute to photon absorption but do not carry enough current to drive the corresponding wire section into a normal-resistive state and trigger an output pulse. To further determine the optimal detector dimensions we pattern such U-shaped NbTiN-nanowires of 90 nm, 75 nm, and 60 nm width over lengths of 25 μm and 40 μm.

| $w_{nw}$(nm) | $l_{nw}$(μm) | $\lambda$(nm) | $\Delta\alpha/\Delta l$ (dB/μm) | $\alpha_{tot}$(%) | $DE_{oc}$(%) | $NEP_{min}$ (W/Hz$^{1/2}$) | $DE_{oc}/(R_{dc}*\delta t)$ |
|---|---|---|---|---|---|---|---|
| 60 | 25 | 1542 | 0.23 | 73.4 | 36.5 | $8.8*10^{-20}$ | $2.8*10^{11}$ |
| 60 | 25 | 768 | 0.57 | 96.2 | 77.3 | $1.9*10^{-20}$ | $6.5*10^{12}$ |
| 60 | 40 | 1542 | 0.23 | 88.0 | 67.7 | $4.0*10^{-20}$ | $8.6*10^{11}$ |
| 75 | 40 | 768 | 0.71 | 99.9 | 80.1 | $2.1*10^{-20}$ | $6.4*10^{12}$ |
| 75 | 40 | 1542 | 0.33 | 95.2 | 52.5 | $7.1*10^{-20}$ | $3.4*10^{11}$ |
| 75 | 25 | 768 | 0.71 | 98.3 | 79.2 | $5.2*10^{-20}$ | $2.0*10^{12}$ |
| 75 | 25 | 1542 | 0.33 | 85.0 | 6.0 | $6.7*10^{-19}$ | $2.6*10^{10}$ |
| 90 | 25 | 768 | 0.91 | 99.5 | 29.7 | $2.6*10^{-19}$ | $1.9*10^{11}$ |
| 90 | 25 | 1542 | 0.47 | 93.3 | 0.4 | $1.1*10^{-17}$ | $1.5*10^{9}$ |

**Table 1.** Detector performance as a function of device dimensions.

Our simulation and measurement results for various nanowire dimensions are summarized in table 1. As expected, in our waveguide geometry high photon absorption rates (>95%) can be achieved for waveguide-coupled visible and infrared photons (26). The results of the measured detection efficiency for a variety of nanowire geometries are shown in figure 2 for both wavelength regimes. Notably, we observe that the measured on-chip detection efficiency ($DE_{oc}$) is strongly correlated with the nanowire width, while the absorption rate rather depends on detector length. This allows us to independently optimize the absorption efficiency ($\alpha_{tot}$) by increasing the nanowire covered length of the waveguide, while maximizing the quantum efficiency ($QE=DE_{oc}/\alpha_{tot}$) by decreasing the wire width.

For visible wavelength photons we observe a sigmoidal dependence of the on-chip detection efficiency on bias current before saturating at 80% in the 60 nm and 75 nm wide nanowires, as shown in Fig. 2a). This behavior is expected from the hotspot model (33,34), which implies that the detection efficiency depends on the hotspot size and reaches the saturation regime when the hotspot diameter approaches the nanowire width. For 60 nm wire width we find that saturation of the detection efficiency sets in already for bias currents as low as 50% of the critical current. Consequently, increasing the nanowire length past 25 μm does not lead to improved detection efficiencies because practically all visible light in the waveguide is already absorbed at this length, see Fig 2a) and table 1.

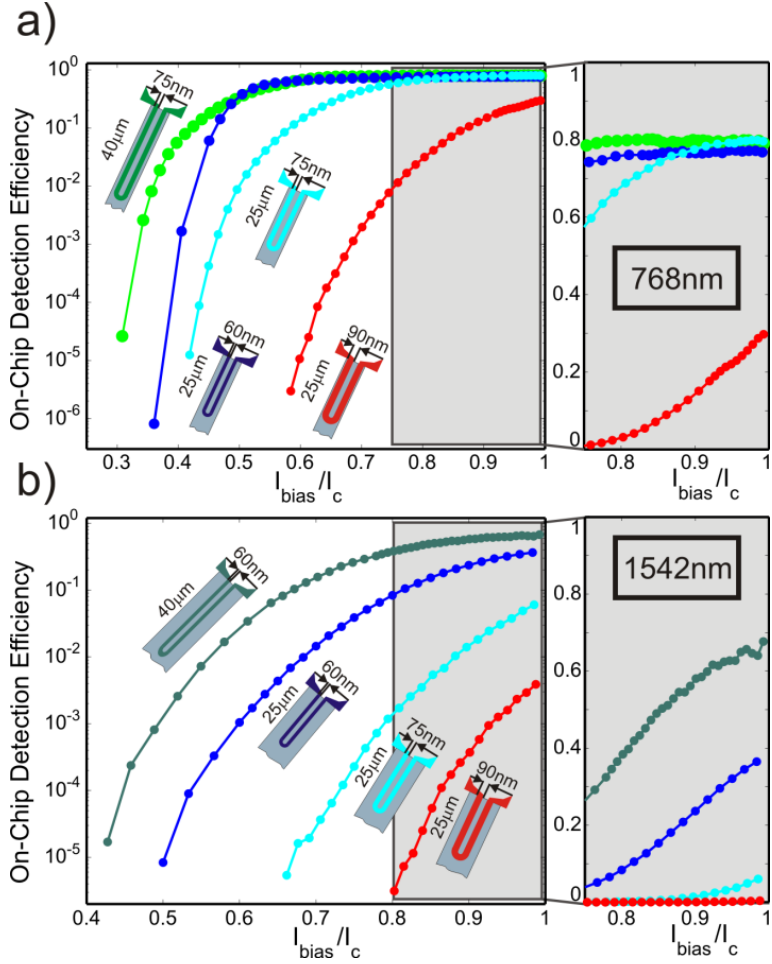

**Figure 2**. On-chip detection efficiency as a function of normalized bias current for NbTiN nanowires of varying lengths and widths; a) for visible (768 nm) photons; b) for infrared (1542 nm) photons. Insets are zoom-in linear plots of detection efficiency in the bias current regime closed to $I_c$.

For telecom wavelength photons the superior performance of narrower nanowires is even more apparent, leading to an increase of the detection efficiency by roughly two orders of magnitude when decreasing the nanowire width from 90 nm to 60 nm, see table 1. Due to the lower photon energy, however, the saturation regime is not quite reached for our SSPDs as the hotspot size and detection efficiency decrease with photon energy, similar to previous observations with telecom wavelength photons (23,34,35). Because the absorption rate of infrared photons is lower than that for visible light (see table 1), it is however possible to achieve another substantial increase in detection efficiency by using longer nanowires. For a 40 μm long SSPD with 60 nm wide nanowires we thus achieve 68% detection efficiency for waveguide-coupled infrared photons.

**Noise Equivalent Power for waveguide-coupled photons.** The Noise equivalent power (NEP) is an important figure of merit for single-photon detector applications including optical time-domain reflectrometry and laser ranging. The NEP critically depends on dark count rate ($R_{dc}$) and detection efficiency (DE), NEP=$h\nu*(2R_{dc})^{1/2}$/ DE, where $h\nu$ is the photon energy. We calculate the $NEP_{oc}$ for waveguide-coupled photons using the on-chip detection efficiency ($DE_{oc}$) data for visible (768 nm) and infrared (1542 nm) photons recorded with 40 μm long SSPDs of 75 nm nanowire width (see Fig 2a) and table 1). In figure 3 we show the corresponding noise equivalent power for the case of minimal stray light ("dark") conditions with the fiber array aligned to the on-chip waveguide ("mounted") or displaced from the waveguide ("unmounted").

The saturation behavior of the detection efficiency for 768 nm photons (see Fig. 2a)) and the low dark count rates observed over the entire bias current range (see Fig. 1 c)) lead to noise equivalent power values in the low $10^{-19}$ W/Hz$^{1/2}$ regime, decreasing to $2.1*10^{-20}$ W/Hz$^{1/2}$ at 58% $I_c$ for visible light. The on-chip detection efficiency for 1542 nm photons on the other hand does not show appreciable saturation behavior but rather decreases monotonically with bias current. For infrared light the corresponding NEP hence reaches a minimum of $7.1*10^{-20}$ W/Hz$^{1/2}$ close to the critical current at 94% $I_c$. Since the dark count rate measurements are very time consuming and limited by stray light rather than material intrinsic mechanisms, we use the data given in figure 1 c) to calculate worst-case NEP values for the other detector geometries summarized in table 1. Note that we observe similar dark count rates for all detectors under test at elevated stray light levels (laboratory under daylight conditions) with shorter data acquisition times though, i.e. no exponential increase at high bias currents. We also note that these NEP values refer to the on-chip detection noise performance. For off-chip applications, the transmission loss at the fiber-to-waveguide interface needs to be taken into account.

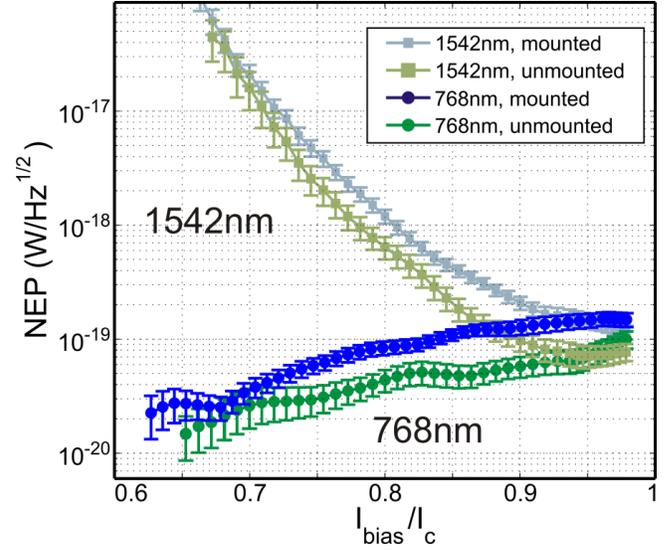

**Figure 3** Noise equivalent power as a function of normalized bias current for visible (768 nm) and infrared (1542 nm) photons under "dark" conditions. See figure 1 for key on stray light levels.

**Single-photon detection with low timing jitter.** For measurements with high timing accuracy and maximal detection efficiency, operation at high bias current is necessary to reduce the detector jitter occurring during the destruction of the superconducting state. High bias currents also result in higher signal-to-noise ratio of the detector output pulses and are hence required for unambiguous pulse discrimination to avoid "false counts" triggered by electrical noise (33). In figure 4 we show the performance of a nanowire SSPD in terms of output pulse amplitude and timing jitter. A representative single shot trace normalized to the maximum of the averaged curve of the SSPD voltage output is shown in Fig. 4a). A fit to the average pulse shape shown in Fig. 4b) (inset) reveals a time constants of $\tau_1$=223 ps for the rising edge and $\tau_2$=1.2 ns for the decay. We thus infer a kinetic inductance of the NbTiN nanowire of $L_k=\tau_2*R_L$=0.75 nH/μm, for an input impedance of $R_L$=50 Ω (36). This value is lower than previous reports on NbN (18)

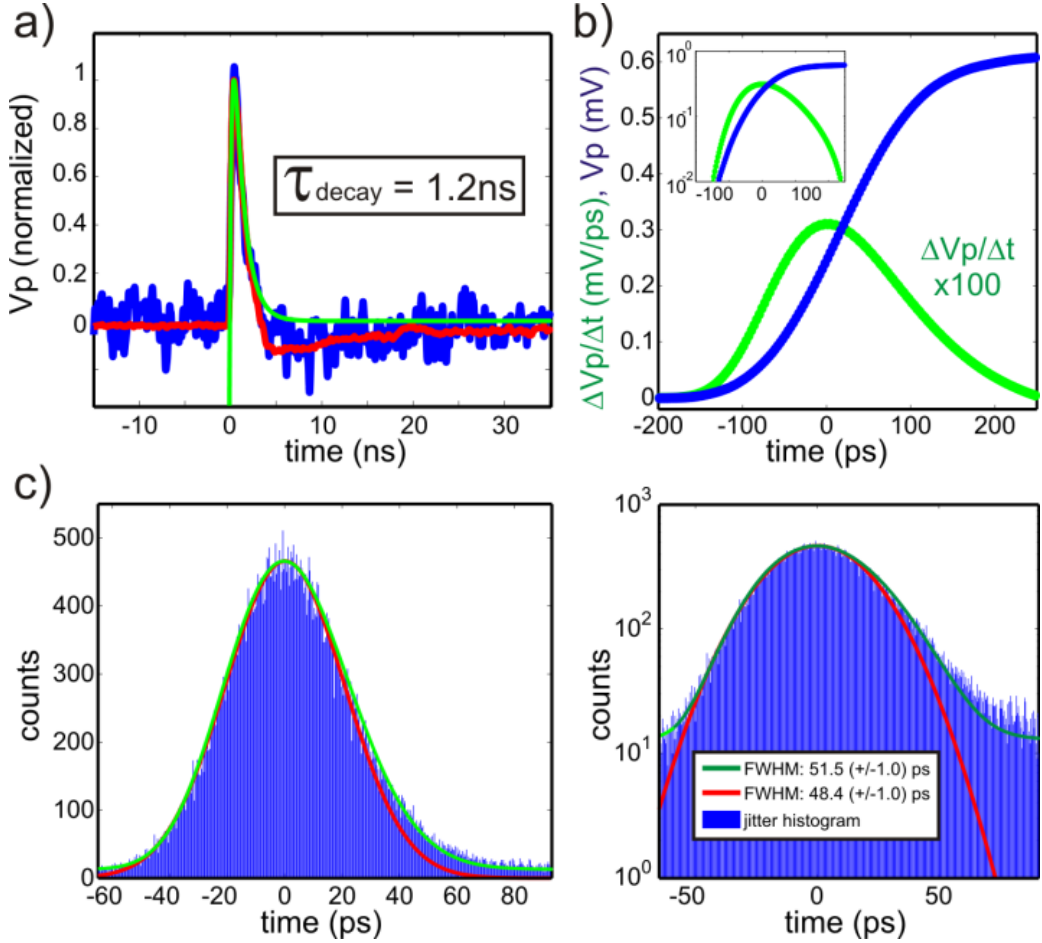

**Figure 4** a) output pulse (normalized) of an NbTiN nanowire SSPD: single shot trace (blue), averaged trace (red) and bi-exponential fit (green) from which we extract a decay time of 1.2 ns. b) asymmetric S-shape of the rising edge of the (averaged) output voltage pulse (0.61 mV before amplification) and its derivative (green, $\Delta V_p/\Delta t$ x100, see inset for logarithmic scale), illustrating the origin of the enhanced contribution to higher pulse jitter values at the trailing part of the pulse distribution; c) histogram of the SSPD output pulse jitter. A Gaussian fit (red) to the data reveals a FWHM jitter value of 48.4 (+/- 1.0) ps. Weighting the Gaussian fit with the derivative of the rising slope of the output pulse (see b)) and accounting for a noise background leads to a slightly increased jitter value of 51.5 ps (green curve).

and implies that GHz counting rates could be achieved by reducing the nanowire length. During the jitter measurements the detector trigger point is set to roughly half the output voltage pulse height where the maximum slope of 3.1 μV/ps (see Fig. 4 b)) is reached. Assuming an RMS noise voltage of 71 μV (see Methods) we thus estimate an instrument limited jitter of 53 ps.

For the experimental determination of the detector jitter we use picosecond laser-pulses and derive a reference trigger signal by splitting off part of the pulses with a fiber optic coupler for detection with a 20 GHz O/E converter. A histogram of the arrival time of the SSPD output pulses relative to the reference signal is shown in figure 4 c). For strongly attenuated laser pulses (<< 1 photon/pulse arriving at the detector) and operation of the SSPD close to the critical current we find the FWHM of the jitter histogram from a Gaussian fit as 48.4 (+/- 1.0) ps. Plotting the jitter histogram distribution on a logarithmic scale (see Fig. 4c)) reveals a slight asymmetry of the histogram envelope with respect to the Gaussian fit. We attribute the increased probability for observing higher time delays of the SSPD output pulse to the asymmetric S-shape of the rising slope of the detectors voltage output. The asymmetry is a consequence of the convolution of the exponentially rising SSPD output pulse due to Joule heating with the fast cooling towards restoring the superconducting state and the instrument bandwidth limitations (6 GHz) of our setup.

While the involved processes require a complex physical model and are not yet fully understood (37,38), we here account for these dynamics by using the derivative of the measured pulse front (see Fig. 4b) as weighting function to the Gaussian fit of the jitter data. Including also a background noise term, we are able to reproduce the envelope of the histogram as shown in Fig. 4c) and obtain a slightly higher SSPD jitter value of $\delta t=51.5$ ps, which is in good agreement with our estimated instrumentation limit.

We note that the jitter value realized here with a waveguide-coupled NbTiN SSPD is lower than those found for most waveguide single-photon detectors (11,21,39) but still somewhat higher than what can be achieved with integrated NbN-on-Si nanowire SSPDs (18). Despite the fact that NbTiN has material properties which make it very attractive for low noise superconducting detector applications, the deposition techniques are not yet as thoroughly investigated as the far more widely used NbN thin films on silicon or sapphire substrates. In particular the stoichiometry of NbTiN is less understood compared to NbN. We hence expect that the thin film quality of NbTiN samples can be further optimized which will result in higher critical current values, in turn leading to higher output pulse amplitudes and a reduced instrument limited jitter value.

**Discussion**

Here we realize low noise single-photon detection over an extended bias current range of our NbTiN-nanowires which allows for operating the SSPDs in a regime where wire defects or constrictions have less influence on the detector performance. Even at bias currents far below the critical current photon counting events can be clearly discriminated from electrical noise without the need for cryogenic amplifier stages due to the intrinsically high signal-to-noise characteristics of the SSPD output pulses. Although challenging in fabrication, engineering the detector dimensions at the nanoscale thus enables us to reach very low noise equivalent power values under realistic measurement conditions. Such low NEP makes these detectors an ideal choice for implementations of photon-counting optical time-domain reflectometers (12) and a broad range of quantum information protocols, particularly in quantum cryptography (6,7).

All three crucial detector features – timing jitter, dark count rate and detection efficiency – can further be combined into one figure-of-merit $DE_{oc}/(R_{dc}*\delta t)$. For the NbTiN-nanowire SSPDs presented here we obtain values in the $10^{11}$-$10^{12}$ regime (see table 1), emphasizing the advantages over other single-photon detector technologies (1) and previous results with NbN-SSPDs on Si-waveguides (18).

In conclusion, we demonstrate waveguide coupled SSPDs with sub-Hz dark count rate all the way up to the critical current where low jitter and high on-chip single-photon detection efficiency are realized. The availability of such low noise detectors, readily integrated with visible and telecom wavelength optical circuitry opens up a broad range of quantum information and nanophotonic applications directly on-chip. Integrating low noise superconducting nanowire single-photon detectors with nonclassical light sources (11,40) and waveguide quantum circuits (41,42) on a complementary metal oxide semiconductor compatible material system shows a promising route to full-scale optical quantum information processing on a monolithic and truly scalable platform.

**Methods**

**Optical and electrical access of the SSPDs.** The sample chip is mounted on a stack of nano-positioners inside a liquid helium cryostat capable of reaching temperatures as low as 1.5 K. Photons from continuous wave (cw) lasers at 768 nm and 1542 nm wavelengths are delivered to the chip via an optical fiber array. Transmission of light from the fiber array into the waveguide is achieved with optical grating couplers patterned into the SiN layer. Different detectors are then optically addressable by positioning the corresponding grating coupler under the fiber array using the low-temperature compatible positioners. The photon flux inside the waveguide is controlled precisely by carefully calibrating the grating coupler transmission and by using calibrated attenuators (18). This calibration procedure is repeated for each device, typical coupling losses are around -11 dB for 1542 nm photons and -13 dB for 768 nm light. The on-chip detection efficiency can then be calculated by dividing the count rate observed at a given bias current value by the photon flux inside the waveguide (26). Considering error contributions from each measurement component and the on-chip photonic devices, the resulting uncertainty of the absolute photon number arriving at the detector is at most 5.9%. For electrical access the nanowires are connected on the chip to gold electrode pads fabricated in an electron beam lithography step using a lift-off technique. These pads are then contacted with an RF-probe to supply the bias current and read out the SSPD photo-response via a 6 GHz bandwidth bias-T. After cooling the devices down to 1.6K we determine the critical current which ranges from 13 μA for the narrowest to 20 μA for the wider nanowires. The SSPD output pulses are amplified and recorded with a single-photon counting system.

**Jitter measurement and pulse characterization.** For the jitter measurement we replace the cw-laser source with a 2.4 ps-pulsed laser (PolarOnyx) and record the output pulses of a 40 μm long, 75 nm wide nanowire SSPD with a high bandwidth oscilloscope (6 GHz, 20 GSa/s) after pulse amplification with high bandwidth (>10 GHz) low noise amplifiers. The shape of the output pulse was empirically fitted with a bi-exponential function $S_p = a(1 - e^{-(t-t_1)/\tau_1}) + be^{-(t-t_2)/\tau_2}$, where the first part accounts for the initial voltage pulse rising sharply at $t_1$ as the nanowire switches to the normal conducting state and the second part describes the exponential decay of the pulse at $t_2$ as the nanowire relaxes back to the superconducting state. Note that the fit shown in Fig. 4a) deviates from the measured data especially at the onset of the pulse and at longer decay times of the falling edge, because the bandwidth limitations of the electrical instrumentation used for the jitter measurements are not taken into account.

To estimate the limiting jitter value of our measurement setup we assume a thermal noise model and calculate the RMS noise voltage $V_{rms}=(4\ k_B T\ \Gamma_{os} R_L)^{1/2}$=70.5 μV at T=300 K using the limiting amplifier bandwidth of the oscilloscope $\Gamma_{os}$=6 GHz and the Boltzmann constant $k_B$. From the difference of critical current (16 μA) and the measured return current (3.8 μA) we obtain the amplitude of the output pulse before amplification as 0.61 mV for an input impedance of 50 Ω. While we assume the SSPD output to be characterized by an exponentially rising voltage pulse for our bi-exponential fit in Fig. 4a), we here determine the slope of the rising pulse edge from the measurement data shown in Fig. 4b).

**Acknowledgments**

We thank Robin Cantor for sample preparation. W.H.P. Pernice acknowledges support by the DFG grant PE 1832/1-1. H.X.T acknowledges support from a Packard Fellowship in Science and Engineering and a CAREER award from the National Science Foundation. We thank Dr. Michael Rooks and Michael Power for their assistance in device fabrication.